# EQUATIONS OF TWO-FLUID HYDRODYNAMICS OF SUPERFLUID HELIUM WITH THE ACCOUNT OF ELECTRIC FIELDS

**V.D. Khodusov, A.S. Naumovets**
*Kharkov V.N. Karazin National University, High-Technology Institute,*
*31 Kurchatov Ave, 61108 Kharkov, Ukraine*
E-mail: khodusov@pht.univer.kharkov.ua

System of two-fluid hydrodynamics of superfluid helium with the account of electric field is obtained. These equations are obtained in kinetic approach using quasi-equilibrium distribution function of quasi-particles, which vanishs collision integral of quasi-particles, and contains dependence on electric field by means of phenomenological parameter $\alpha$. Using experimental data at temperature range of 1,4 - 2 K, where basic role plays roton hydrodynamics, the value of phenomenological parameter, is obtained.
**KEY WORDS**: superfluid Helium, equations of two-fluid hydrodynamics, second sound, rotons
**PACS**: 67.40. Pm, 67.40. Bz.

In works [1, 2] was experimentally studied the electrical response caused by second sound waves (SSW) or by torsion vibrations in superfluid helium. This response is provided by the existence of dielectric polarization in the absence of electric field. Observation of electric field by thermal excitation of SSW and actually the fact of resonant excitation of these waves by the variable electric field, the resonance curve coincidence for electric and temperature fields suggests these fields to take part at the same oscillating process and linearly banded with each other. In addition the resonant absorption of electromagnetic waves has been experimentally revealed in superfluid helium [3-5]. All these experiments were set in the range of sufficiently low temperatures where the description of superfluid helium state can be carried out in the frames of two-fluid hydrodynamics. Attempts of a theoretical explanation of these experiments have been made in works [6-11]. For these experiments theoretical explanation nessesary exists of receiving of two-fluid hydrodynamics equations with the account of electric fields.

On basis of Bose quasi-particles gas dynamics general theory with variable parameters [12] one can get coupled equations that describe the behavior of medium and quasi-particles. In this way Landau two-fluid hydrodynamics equations are deduced in works [12, 13]. When getting these equations a significant part plays concept of quasi-local distribution function that turns to zero collision integral and include electric field, which can be taken into account by the instrumentality of phenomenological constant.

## KINETICS OF BOSE QUASI-PARTICLES WITH THE ACCOUNT OF ELECTRIC FIELD.

There were obtained equations of gas dynamics of Bose quasiparticles in the account of external fields. Based on these equations, a system of equations two-fluid hydrodynamics of superfluid helium was obtained. We obtain the same method, these equations taking into account the electric field. In the kinetic theory the state of quasi-particles gas is characterized by quasi-particles distribution function $N \equiv N(\mathbf{p}, \mathbf{r}, t)$, which satisfies the kinetic equation that has a form of Boltzmann equation, where $\mathbf{p} = \hbar \mathbf{k}$ – the momentum of quasi-particles.

$$\left( \frac{\partial}{\partial t} + \mathbf{g} \frac{\partial}{\partial \mathbf{r}} - \frac{\partial \varepsilon}{\partial \mathbf{r}} \frac{\partial}{\partial \mathbf{p}} \right) N = \left( \dot{N} \right)_{coll} \qquad (1)$$

here $\mathbf{g} \equiv \partial \varepsilon / \partial \mathbf{p}$ – the group velocity of quasiparticles; $\varepsilon \equiv \varepsilon(\mathbf{p},\mathbf{r},t)$ – the Hamiltonian of the quasiparticles, which coincides with its local energy; $\left( \dot{N} \right)_{cm}$ – the collision integral of quasiparticles, which takes into account the processes of collision, merger, dissolution and emission of quasiparticles. This equation was first used by the A.I. Akhiezer in [14] to describe the nonequilibrium state of the system of phonons in crystals.

Explicit dependence of the energy of quasiparticles on the spatial $(\mathbf{r})$ and temporal (t) variables is due to its dependence on external fields in terms of their low spatial heterogeneity and slow changes in time (when the condition of adiabaticity changes in these fields), i.e. when the scale of the inhomogeneity of the fields $L_c$ significantly larger than the characteristic wavelengths of quasiparticles $\lambda$, and the characteristic times of changes in these fields $t_c$ much larger than the characteristic periods of oscillations of quasiparticles $T_{кв}$.

$$L_c \gg \lambda, \quad t_c \gg T_{кв} \qquad (2)$$

In the absence of external fields the solution of the kinetic equation (1), vanishing collision integral, is the Planck equilibrium distribution function:

$$\bar{N} = \left( \exp \frac{\varepsilon_0}{k_B T_0} - 1 \right)^{-1} \qquad (3)$$

where $T_0$ – equilibrium temperature.

We assume later that in the quickly relaxing system of quasiparticles interaction processes at which the energy and momentum is conserved (N-protcesses) are determine. Non-conservation of momentum and energy due to the interaction of the quasiparticles, such as particles, leading to their decay, the quasiparticle scattering by impurities, defects, etc. We will denote by $\tau_N$ the characteristic time of quasi-particle interaction due to N-processes, and by $\tau_R$ - characteristic time of quasi-particle interaction for through processes in which the total momentum is not conserved. Then the condition that the normal processes are decisive, can be written as follows: $\tau_N \ll \tau_R$. Such a situation occurs in the pure crystals and quantum liquids at low temperatures [12, 13], in particular, these conditions are satisfied in the experiments in [1 - 5].

Excited state of HeII is described by the gas of quasi-particles (phonons, rotons). The carried experiments [1-5] denote that vectors of electric field **E** and electric induction **D** (polarization vector) enter into the set of thermodynamic parameters that characterize the properties of superfluid helium. That is why natural to assume that the quasiparticle energy depends linearly on the electric field [7]. The momentum of the quasiparticle **p** is the only vector characterizing the quasiparticle and therefore it is necessary to minimize the electric field vector with it to recieve a scalar. Because of these considerations, quasiparticles energy in He II in the presence of superfluid speeds $\mathbf{v}_s$ can be written as follows

$$\varepsilon(\mathbf{p}, \mathbf{E}) = \varepsilon(\mathbf{p}) + \mathbf{p}(\alpha \mathbf{E} + \mathbf{v}_s) \qquad (4)$$

So as to perform the property of quasi-particle energy invariance with respect to time inversion it is necessary for $\alpha$ to satisfy the condition $\alpha(-t) = -\alpha(t)$. Confirmation of this property will be seen from the further relations (24). Recent experiments on the splitting of the resonant line of the microwave absorption on the roton frequency in electric fields evidence in favor of this hypothesis. Furthermore, in [15] was calculated dipole moment of the roton, which is proportional to the momentum of the roton.

It is significant to note that in present work we do not assume the presence of spontaneous polarization of fluid in superfluid state. Though the terms that contain electric field in quasi-particle energy have the form of dipole energy in external field what may mean the appearance of superfluid liquid unit volume polarization. However ascertainment of these circumstances calls construction of a theory of current effect in microscopic approach, which goes beyond the scope of current work.

If in some instant of time the system of quasi-particles goes out of its equilibrium state, then during relaxation time of normal processes quasi-local balance establishes and it is characterized by distribution function $N_0$, which vanish collision integral, and have a form [12]:

$$N_0 = \left( \exp \frac{\varepsilon - \mathbf{p}(\mathbf{w} - \alpha \mathbf{E})}{k_B T_0 (1 + \theta)} - 1 \right)^{-1} \qquad (5)$$

where $\mathbf{w} = \mathbf{v}_n - \mathbf{v}_s$ - relative velocity, $\mathbf{v}_n$ - normal velocity, $\mathbf{v}_s$ - superfluid velocity, $\theta = (T - T_0)/T_0$ - is relative temperature, **E** – electric field, $\varepsilon(\rho, \mathbf{p}) = \varepsilon_0(\rho_0, \mathbf{p}) + \frac{\delta \varepsilon}{\delta \rho} \delta \rho$.

Quasi-equilibrium distribution function is associated with the thermodynamically equilibrium distribution function of quasiparticles with an accuracy to quadratic terms in small quantities, by the relation:

$$N_0 = \bar{N} - \frac{\bar{N}(\bar{N}+1)}{k_B T_0} \left( \frac{\partial \varepsilon}{\partial \rho} \delta \rho - \mathbf{p}(\mathbf{w} - \alpha \mathbf{E}) - \varepsilon_0 \theta \right) \qquad (6)$$

In sub-equilibrium statistical state, the solution of Boltzmann equation in the hydrodynamic approximation is sought in the form:

$$N = N_0 + \delta N, \quad (|\delta N| \ll N_0)$$

where $N_0$ – the local-equilibrium distribution function (5), depending on the hydrodynamic quantities, and $\delta N$ depends also on their gradients. Linearized Boltzmann kinetic equation will be:

$$\left( \frac{\partial}{\partial t} + \tilde{\mathbf{g}} \frac{\partial}{\partial \mathbf{r}} - \frac{\partial \tilde{\varepsilon}}{\partial \mathbf{r}} \frac{\partial}{\partial \mathbf{p}} \right) N_0 = \left( \delta \dot{N} \right)_{coll} \qquad (7)$$

where $\tilde{\mathbf{g}} \equiv \partial \tilde{\varepsilon}/\partial \mathbf{p}$, $\tilde{\varepsilon} \equiv \varepsilon(\rho, \mathbf{p}) - \mathbf{p}(\mathbf{w} - \alpha \mathbf{E})$. After multiplying equation (7) by momentum **p** and energy $\varepsilon$ of quasi-particles with subsequent momentum integrating we will obtain gas-dynamic system of equations. Before writing gas-dynamic system of equations we will derive thermodynamic quantities entering in it that characterize locally equilibrium state.

## THERMODYNAMICS OF BOSE QUASI-PARTICLES WITH THE ACCOUNT OF ELECTRIC FIELD.

Following the work [12], we introduce density of thermodynamic potential of quasi-particles $F_0$ by expression:

$$F_0 = -k_B T \int d\tau_\mathbf{p} \ln(1+N_0) \tag{8}$$

where $\int d\tau_\mathbf{p} ... = \dfrac{1}{(2\pi\hbar)^3} \int d\mathbf{p}...$, and $F_0$ is banded with density of internal energy as

$$F_0 = U - S_0 T - \mathbf{j_0}\mathbf{v}_n \tag{9}$$

For thermodynamic potential next thermodynamic identities is true:

$$dF_0 = -S_0 dT - \mathbf{j_0} d\mathbf{w} + \tilde{\zeta} d\rho - \frac{\mathbf{D_0}}{4\pi} d\mathbf{E}$$

$$dU = TdS_0 + \mathbf{v}_n d\mathbf{j_0} + \tilde{\zeta} d\rho + \mathbf{j_0} d\mathbf{v}_s - \frac{\mathbf{D_0}}{4\pi} d\mathbf{E} \tag{10}$$

If we know $F_0$ as the function of $T$, $\mathbf{w}$, $\rho$, $\mathbf{E}$ we will find the momentum density of quasiparticles $\mathbf{j_0}$, the heatcapacity C and the entropy S of quasiparticle, $\mathbf{D}_0$, the density of the quasiparticles (normal density) $\rho_n$, chemical potential $\tilde{\zeta}$:

$$\mathbf{j_0} = -\left(\frac{\partial F_0}{\partial \mathbf{w}}\right)_{\rho,T,\mathbf{E}} = \int d\tau_\mathbf{p} N_0 \mathbf{p} = \rho_n(\mathbf{w}-\alpha\mathbf{E}), \qquad \rho_n = \frac{\partial \mathbf{j_0}}{\partial \mathbf{v}_n} = -\left(\frac{\partial^2 F_0}{\partial \mathbf{v}_n^2}\right)_{\rho,T,\mathbf{E}}$$

$$S_0 = -\left(\frac{\partial F_0}{\partial T}\right)_{\rho,\mathbf{w},\mathbf{E}} = \int d\tau_\mathbf{p}\left\{k_B \ln(1+N_0) + N_0 \frac{\varepsilon - \mathbf{p}\mathbf{v_n}}{T}\right\}, \qquad C = T\left(\frac{\partial S_0}{\partial T}\right)_{\rho,\mathbf{w},\mathbf{E}}$$

$$\frac{\mathbf{D}_0}{4\pi} = -\left(\frac{\partial F_0}{\partial \mathbf{E}}\right)_{\rho,T,\mathbf{w}} = -\alpha\mathbf{j_0}, \qquad \tilde{\zeta} = -\left(\frac{\partial F_0}{\partial \rho}\right)_{\mathbf{w},T,\mathbf{E}} = \int d\tau_\mathbf{p} N_0 \frac{\partial \varepsilon}{\partial \rho} \tag{11}$$

We take into account the influence of environment by changing to the thermodynamic potential $F = F_0 + U_0(\rho,\mathbf{E})$. Here $U_0$ – the internal energy of the system for $T=0$, satisfying the thermodynamic identity:

$$dU_0 = \zeta(0) d\rho - \frac{\mathbf{D}(0)}{4\pi} d\mathbf{E} \tag{12}$$

where, $\zeta(0)$, $\mathbf{D}(0)$ - the chemical potential and the vector of electric induction respectively, at $T=0$. With this in mind we obtain:

$$dF = -S_0 dT - \mathbf{j_0} d\mathbf{w} + \left(\zeta(0)+\tilde{\zeta}\right) d\rho + \left(\alpha \mathbf{j_0} - \frac{\mathbf{D}(0)}{4\pi}\right) d\mathbf{E} \tag{13}$$

It is easy to obtain the following relation between thermodynamic variables $p$, $\zeta = \zeta(0)+\tilde{\zeta}$ and $F$:

$$P = \zeta\rho - F \tag{14}$$

Note that the following relation takes place: $\mathbf{D}(0) = \varepsilon(0)\mathbf{E}$, at $T=0$; where $\varepsilon(0)$ – the dielectric constant.

## EQUATIONS OF TWO-FLUID HYDRODYNAMICS WITH THE ACCOUNT OF ELECTRIC FIELDS.

Applying the standard procedure [13, 14] from the kinetic equation we can get hydrodynamic system of equations:

$$\begin{cases} \dfrac{\partial}{\partial t} j_{0i} = -\dfrac{\partial}{\partial x_l}(j_{0i}v_{nl} - F_0 \delta_{il}) - \tilde{\zeta}\dfrac{\partial \rho}{\partial x_i} - j_{0l}\dfrac{\partial}{\partial x_i}(v_{sl} + \alpha E_l) \\ \dfrac{\partial U}{\partial t} = -div\left[(TS_0 - \mathbf{j_0}\mathbf{v}_n)\mathbf{v}_n\right] + \tilde{\zeta}\dfrac{\partial}{\partial t}\rho + \mathbf{j_0}\dfrac{\partial}{\partial t}(\mathbf{v}_s + \alpha\mathbf{E}) \end{cases} \tag{15}$$

Clearly, this system of equations is not confined. It is nessesary to add the equations for the density $\rho$, $\mathbf{V}_s$ and the electric field $\mathbf{E}$. One of these equations is the equation of conservation of mass, i.e. the continuity equation:

$$\dot{\rho} + div\mathbf{j} = 0 \tag{16}$$

where $\mathbf{j}$ - full momentum of the system defined by the following expression:

$$\mathbf{j} = \mathbf{j}_0 + \rho\mathbf{v}_s = \rho_n(\mathbf{v_n} - \alpha\mathbf{E}) + \rho_s\mathbf{v}_s \tag{17}$$

$\rho_s = \rho - \rho_n$ - density of the superfluid component.

We find the other equations from the law of conservation of total momentum. After differentiating $\mathbf{j}$ with respect to time and excluding $\dot{\rho}$ and $\dot{\mathbf{j}}_0$ with the help of equations (15) and (16), we obtain:

$$\frac{\partial}{\partial t}j_i = -\frac{\partial}{\partial x_l}\left[\rho_n v_{ni}v_{nl} + \rho_s v_{si}v_{sl} - \alpha\rho_n(v_{si}E_l + v_{sl}E_i) - (F - \zeta\rho)\delta_{il} - \left(\frac{\varepsilon_0}{4\pi} + \alpha^2\rho_n\right)E^2\delta_{il}\right] -$$

$$-[\mathbf{j}, rot\mathbf{v}_s] + \rho\left[\frac{\partial v_{si}}{\partial t} + \nabla_i\left(\tilde{\zeta} - \frac{\mathbf{v}_s^2}{2}\right)\right] \tag{18}$$

As follows from (18), the law of conservation of momentum will be satisfied if:

$$[\mathbf{j}, rot\mathbf{v}_s] = 0 \tag{19}$$

$$\frac{\partial v_{si}}{\partial t} + \nabla_i\left(\zeta - \frac{\mathbf{v}_s^2}{2}\right) = 0 \tag{20}$$

If we use the relation (14), equation (18) is written as:

$$\frac{\partial}{\partial t}j_i = -\frac{\partial}{\partial x_l}\left[\rho_n v_{ni}v_{nl} + \rho_s v_{si}v_{sl} - \alpha\rho_n(v_{si}E_l + v_{sl}E_i) - P\delta_{il} - \left(\frac{\varepsilon_0}{4\pi} + \alpha^2\rho_n\right)E^2\delta_{il}\right] \tag{21}$$

For obtaining this system of equations we have used Maxwell's equations for longitudinal fields:

$$\frac{\partial \mathbf{D}}{\partial t} = 0, \quad div\mathbf{D} = 0 \tag{22}$$

Equation (16), (18) - (22) is a complete system of equations of two-fluid hydrodynamics with an electric field.

To determine the phenomenological parameters $\alpha$ we use the linearized system of equations two-fluid hydrodynamics:

$$\begin{cases} \frac{\partial}{\partial t}\mathbf{j} = -\nabla P; \quad \dot{\rho} + div\mathbf{j} = 0; \\ \frac{\partial}{\partial t}\mathbf{v}_s + \nabla\zeta = 0; \quad \frac{\partial}{\partial t}S_0 + \overline{S}\cdot div\mathbf{v}_n = 0; \\ \frac{\partial(\varepsilon(0)\mathbf{E} - 4\pi\alpha\mathbf{j}_0)}{\partial t} = 0; \quad div(\varepsilon(0)\mathbf{E} - 4\pi\alpha\mathbf{j}_0) = 0. \end{cases} \tag{23}$$

From system (23) we obtain the relations:

$$\frac{\partial \mathbf{E}}{\partial t} = -\frac{4\pi\alpha\overline{S}}{\varepsilon_0}\nabla T = \frac{4\pi\alpha\rho_n}{\varepsilon_0 + 4\pi\alpha^2\rho_n}\cdot\frac{\partial \mathbf{w}}{\partial t}$$

$$\frac{\partial \mathbf{j}_0}{\partial t} = -S\nabla T \tag{24}$$

from which it follows that $\alpha(-t) = -\alpha(t)$, i.e. quasi-particle energy is invariance with respect to time inversion.

Excluding from the first equations velocities $\mathbf{v_n}$, $\mathbf{v_s}$ and chemical potential we get ordinary equations that describe propagation of first and second sound:

$$\frac{\partial^2 \rho}{\partial t^2} - \Delta P = 0, \quad \frac{\partial^2 S_0}{\partial t^2} = \frac{\rho_s}{\rho_n}S_0^2\Delta T + \frac{S_0}{\rho}\Delta P \tag{25}$$

Changing to the Fourier components in equation (24) and introducing the wave phase velocity by the relation $u = \omega/k$ and considering the longitudinal electric field, we obtain:

$$\varepsilon(0)E = \frac{4\pi \overline{S} \alpha}{u} T' \qquad (26)$$

We will define a phenomenological parameter $\alpha$, using the experimental data of [1]. In this study, a longitudinal electric field was used and took into account its influence on the propagation of second sound. From the analysis of experimental data yielded the following relationship between the fluctuation of temperature $T'$ and electric field potential $\Delta U$ in SSW: $\frac{T'}{\Delta U} \approx \frac{2e}{k_B}$. Because of the geometry of the resonator connection of the electric field with the potential is linear: $E = \Delta U / d$, where $d = 2,8\ cm$ - the length of the resonator. Using equation (26), assuming $\varepsilon(0) \approx 1$, we get:

$$\alpha \approx \frac{u_2}{S_0}\frac{E}{T'} = \frac{u_2 d}{S_0}\frac{k_B}{2e} \qquad (27)$$

From the expression (27) follows that the parameter $\alpha$ depends on temperature through the temperature dependence of entropy $\overline{S}$ and the velocity of second sound.

We will give an estimate of parameter $\alpha$ at $T = 1,4K$ in which the velocity of second sound is equal to $u_2 \approx 19,7 m/\sec$ and entropy $S_0 = 0,132 J/g \cdot K$: $\alpha \approx 6,91 \cdot 10^{-9} \frac{sm}{c \cdot sgs_q}$.

## CONCLUSION

Received system of equations (23) makes it possible to describe the experimental results [1-5], which were carried out at temperatures of 1,4 ÷ 2 K. In this temperature range roton gas acts the main part in thermodynamic and kinetic properties of HeII. At the expense of fast roton-roton interactions sets hydrodynamic regime that is described by two-fluid hydrodynamics with account of electric fields. This system of equations can be obtained in the kinetic approach, define the dependence of local equilibrium distribution function by the electric field using a phenomenological parameter $\alpha$.

## ACKNOWLADGMENTS

We are grateful to I.N. Adamenko, Yu.A. Kirochkin, A.S. Rybalko for fruitful discussion.